\documentclass[11pt]{article}
\usepackage{doi}
\usepackage{a4wide}
\usepackage[T1]{fontenc}
\usepackage{graphicx}
\usepackage[authoryear]{natbib}
\usepackage{soul}
\usepackage{amsmath}
\usepackage[group-separator={,}]{siunitx}
\usepackage{caption}
\usepackage{subcaption}
\usepackage{multirow}
\usepackage{multicol}
\usepackage{placeins}
\usepackage{balance}
\usepackage{booktabs}
\usepackage{makecell}
\usepackage{amssymb}
\usepackage{xcolor}

\begin{document}
%
\title{Performance Consistency of Learning Methods for Information Retrieval Tasks}


\author{Meng Yuan, Justin Zobel \\
University of Melbourne \\
Melbourne, Australia}

%

%
\maketitle              
\begin{abstract}
A range of approaches have been proposed
for estimating the accuracy or robustness of the measured performance of IR methods.
One is to use
bootstrapping of test sets, which, as we confirm, provides an estimate of variation in
performance.
For IR methods that rely on a seed, such as those that involve
machine learning, another approach is to use a random set of seeds
to examine performance variation.
%
Using three different IR tasks
we have used such randomness to examine a range of traditional statistical learning models
and transformer-based learning models.
While the statistical models are stable, the transformer models show huge variation
as seeds are changed.
In 9 of 11 cases the F1-scores (in the
range 0.0--1.0) had a standard deviation of over 0.075; while 7 of 11 precision values
(also in the range 0.0--1.0) had a standard deviation of over 0.125.
This is in a context where differences of less than 0.02 have been used
as evidence of method improvement.
Our findings highlight the vulnerability of transformer models to training instabilities
and moreover raise  questions
about the reliability of previous results, thus
underscoring the need for rigorous evaluation practices.
\end{abstract}

%
\section{Introduction}

In information retrieval (IR), a standard approach to evaluation of a method is to apply
the method to a set of test data and report a numerical score.
A simple example of a task where this kind of approach is used is binary text classification.
If the method is based on some form of machine learning
(a terminology used here in a broad sense, encompassing transformers as well
as statistical methods),
training data is used to establish the classifier, which
is then applied to test data, producing a set of outputs that are either true or false.
These outputs are then translated to measures such as precision or the F1-score.

A typical element of such evaluations is that the score is reported to three or four
decimal places without consideration given to the possibility of uncertainty, as
might be measured by confidence or deviation.
However, uncertainty can arise in a range of ways.
One is the selection of test data.
Statistical methods such as bootstrapping provide a way to estimate score variability
arising from composition of the test data, and
have been occasionally (if infrequently) used in such evaluations.
Here we report bootstrap results on three classification tasks and nine ML methods,
confirming its value as an indicator of accuracy.

Another source of uncertainty for ML methods is the choice of seed,
which influences multiple aspects 
of model training including the ordering of training data, 
initialization of parameters, and update order.
This uncertainty is the primary focus of our work.

In statistical machine learning, such as linear 
regression, training is fully deterministic: every use of a random 
seed can be traced, and the entire process can be reconstructed to 
understand precisely how parameters were updated. 
However, this determinism does not hold in deep learning.
Even small differences in factors such as ordering of training data
can in principle yield substantial performance variations.
That is, the training process itself may to some extent be unstable,
but it is rare for papers to acknowledge this instability, let alone test or
report it \citep{zhuang2022mlsy}.

Such instability undermines reproducibility and indeed the usability of
the methods.
It also means that, without statistical examination or acknowledgement of variability, research
results on these methods are not reliable -- particularly as researchers in some cases
choose random seeds to optimise performance.
These concerns are not new; they have been examined for
specific methods in several prior works
\citep{antoniak2021acl,dutta2022,Fellicious2020mlods,isensee2024nnunet,schader2024epd}.
However, to our knowledge these issues have not been previously examined across a
basket of ML methods in the context of IR tasks.

We examine how seed choices affect model performance for nine statistical ML and
trans\-former-based approaches on three representative IR tasks:
sentiment analysis, fake news detection, and binary question answering.
The statistical methods yield consistent results but the transformer-based methods
show such high levels of variation that in many cases the `true' results are
effectively unknown to even a single decimal place.
These results demonstrate that it is critical that researchers adopt more robust research practices.


\section{Related work}

A commonly used approach for evaluation of the robustness of algorithms and metrics is bootstrapping,
a statistical method for assessing the variance in a data set that has the benefit of not
relying on distributional assumptions.
The principle of bootstrapping is that
a dataset of $n$ items is sampled $n$ times with replacement,
creating a new dataset that is consistent in class distribution with the original data.

Test set bootstrapping has been used for IR metric evaluation.
For example,
\citet{sakai2006sigir} has used bootstrapping to estimate the minimum performance gap required for reliable discrimination between systems.
\citet{Smucker2007cikm} then adopted \citeauthor{sakai2006sigir}'s approach to
test three statistical methods for evaluating measure sensitivity: bootstrapping,
permutation, and the t-test.
Later work further explored bootstrap-based confidence intervals for
rank correlation among metrics \citep{sakai2007ipsj} and studied evaluation stability
under different sampling and pooling strategies involving bootstrapping \citep{urbano2013sigir}.
Collectively, these studies establish bootstrapping as a principled approach for assessing both the discriminative power of IR evaluation measures and the robustness of experimental conclusions in~IR.

The use of random seeds for evaluation of ML variation is more diverse, in part because
random seeds are involved in a variety of stages in model training and evaluation.
For example, they help ML methods to avoid falling into local optima and therefore
help avoid bias on model performance \citep{goodfellow2016mitpress}. 
Random seeds have been examined in many previous works on ML evaluation,
and have been considered as a common approach for robustness testing of machine learning algorithms \citep{reimers2017emnlp, henderson2018aaai, dodge2019emnlp}. 
Over time
this became a convention in the evaluation process and many recent works do not report
the variance of performance with respect to multiple random seeds;
instead, only an average is given.

Several recent papers have explored concerns about the randomness involved in
model training, especially for deep learning models \citep{Fellicious2020mlods, dutta2022, antoniak2021acl, schader2024epd}.
The training process of deep learning models is to some extent non-deterministic due to a
variety of components including the internal layers and GPU optimisation \citep{zhuang2022mlsy};
these components contribute to the randomness of the training process along with the random seed itself.
%
Although quantitative studies on effects of random seeds in deep learning models are limited, a couple of recent studies have explored them from a variety of perspectives.
\citet{Fellicious2020mlods} examined the performance of ANN models with three optimisers using different random seeds and observed significant variance in the performance of the models, ranging from 77\% to 83\%.
However, a limitation of this experiment is that they only examined a single ANN model with different optimisers and the performance variation within ANN model may not generalise to other deep-learning models or tasks.
\citet{dutta2022} undertook a large-scale study of the effect
of seeds with 114 machine learning projects collected from Github, and identified over 400
tests that failed due to change of seeds.
The majority of these test failures remain unresolved.

Other work seeks to achieve reproducibility when training with the same machine while fixing the
seed \citep{hill2024ESM, schader2024epd}.
Their solution successfully reproduced identical results by fixing the random seed
and a series of dependent components of the models that involve randomisation,
the same results cannot be reproduced on a different machine.

Similar concerns have also been raised in other areas of deep learning. 
A recent study in 3D medical image segmentation found that many 
reported gains of new architectures are not observed when tested with 
strong baselines and rigorous validation protocols, thus providing evidence
of the need for 
higher standards of evaluation across the field~\citep{isensee2024nnunet}. 
This observation is closely related to the issue we examine here: 
reported performance can be strongly influenced by overlooked factors, 
in our case random seed choices, which call into question the 
robustness of widely adopted evaluation practices.

\section{Methods}

We examine the precision of evaluation on three tasks,
sentiment analysis, fake news detection, and binary Q\&A. 
These tasks are described below, but we first describe our evaluation strategy.

\subsection*{Evaluation of accuracy}

One approach to evaluation of accuracy that we used was bootstrapping of the test data,
or equivalently of the per-item test results.
Use of bootstrapping provides a control for the later experiments as it indicates the natural
variability in the outcomes; if randomness influences the results then it is reasonable for the
variation to be consistent with that inherent in the data.
For each task and ML method, we generated 100 bootstrap samples of the outcomes and computed
the mean and standard deviation.

The other approach that we used was to generate sets of random seeds.
Such seeds are involved in many components of training and fine-tuning of ML models,
noting that, for traditional models such as logistic regression or SVMs, once the
training set is fixed the training process is deterministic and the effect of the seed is negligible.
However, random forests are an exception, as their feature subsampling and bootstrapping steps
depend on the seed.
(These and other ML methods are introduced below.)
In contrast, for deep learning models such as BERT and RoBERTa, the seed influences multiple parts of the training pipeline.
These include initialisation of model weights, the order of training instances in each epoch, dropout operations during training, and optimiser updates that are further subject to GPU-level scheduling.
Together these factors mean that,
even with the same configuration, two training runs with different random seeds can lead to substantially different model outcomes.
Table~\ref{tab:seed_influence} summarises the specific ways in which random seeds influence the training process for the types of models used in our experiments.

\begin{table}[t]
\centering
\caption{Influence of random seeds on different types of ML model.}
\label{tab:seed_influence}
\begin{tabular}{@{~~}m{3.1cm}m{3.1cm}m{4.1cm}}
\toprule
Model type & Example models & Components influenced \\
\midrule

Machine-learning models & 
\begin{minipage}{3.1cm}
  Logistic regression \\ SVM \\ Naïve Bayes \\ Random forest \\
\end{minipage}
 &
\begin{minipage}{4.3cm}
  Training data shuffling \\ Feature subsampling \\
\end{minipage} \\
\midrule

Encoder-based transformers & 
\begin{minipage}{2.8cm}
  BERT \\ RoBERTa \\ XLNet \\ 
\end{minipage}
 &
\begin{minipage}{4.3cm}
  Weight initialisation \\ Dropout layers \\ Training data order \\ GPU scheduling \\
\end{minipage} \\
\midrule

Encoder--decoder transformers & 
\begin{minipage}{2.8cm}
  BART \\ T5 \\
\end{minipage}
 &
\begin{minipage}{4.3cm}
  Weight initialisation \\ Dropout layers \\ Teacher forcing randomness \\ Optimiser update order \\
\end{minipage} \\
\bottomrule

\end{tabular}
\end{table}

To evaluate variation under random seeds,
we generate 100 seeds using the Python \texttt{random} function.
Each seed is then used to set the \texttt{random\_state} or \texttt{random\_seed} argument
where available in the model implementation.
For each model on each task, training is fully repeated for each of the 100 seeds.
This allows us to directly examine the degree to which random seed variation influences model performance, independent of other factors.

\subsection*{Tasks}

We used three IR tasks to explore the variation in performance of nine different ML models;
all of these have previously been addressed using these ML models or other similar techniques.
In each case we use the same pre-processing methods,
to ensure that comparisons are fair and consistent.
Note that not all models could be used on all tasks, as illustrated in
Table~\ref{tab:models-summary}.
The tasks are as follows; the related datasets are described in the Experiments section.

\subsubsection*{Sentiment analysis.}
This is a text-based classification task that involves identification
and classification of opinions, emotions, or attitudes expressed within textual data.
The primary objective is to categorise such sentiment in a given text as
positive, negative, or neutral,
or in some cases into more specific emotional states such as joy, sadness, anger, or fear.

Applications of sentiment analysis include analysis of customer feedback,
monitoring of social media,
market research for brand perception,
and development of recommender systems tailored to user preferences \citep{pang2008opinion,liu2012sentiment}.
Common methodologies include lexicon-based approaches,
which rely on predefined sentiment dictionaries \citep{taboada2011lexicon};
machine-learning classifiers such as na\"ive Bayes,
SVMs, and logistic regression \citep{pang2002thumbs};
and deep-learning techniques based on neural-network architectures such as CNNs and RNNs \citep{kim2014convolutional, tang2015document}, LSTMs \citep{hochreiter1997long}, and transformers such as BERT and RoBERTa~\citep{devlin2019bert, liu2019roberta}.

\subsubsection*{Fake news detection}
This is a classification task in which
the aim is to determine whether a given claim is truthful or fabricated.
It typically involves identifying misleading or deceptive content
that mimics legitimate news.

Commonly adopted approaches to fake news detection include:
content-based approaches, where linguistic features, writing style,
and sentiment analysis are used to detect inconsistencies existing in the
text \citep{rashkin2017emnlp, perez2018iccl};
contextualised approaches, where external sources of information are used
for fact-checking and validation of claims \citep{popat2018emnlp, thorne2018naaclhlt};
and network-based approaches, where sources and patterns of news publication
are analysed to detect abnormal behaviour that is characteristic of
rumour-spreading \citep{shu2019wsdm, vosoughi2018science}.
Here, we assess both machine-learning models and deep-learning models on
their ability to identify fake news based only on the contents of the item,
without reference to external sources of information.

\subsubsection*{Binary Q\&A}
This is a fact-checking task that takes a statement in natural language form
and returns {\emph{yes}} or {\emph{no}} \citep{BoolQ2019naacl, Wang2019NeurIPS, Wang2019ICLR}.
The challenge of this task is to learn the relationship between the input statements
and the binary labels based on factual correctness rather than semantic similarity.
Binary Q\&A is regarded as more challenging that the tasks above as it requires models
with complex structure and large numbers of parameters.
We use two types of models to examine their robustness on this task:
encoder-based models
and encoder-decoder-based models.
These are discussed below.

\section{Experiments}

\subsubsection*{Data}
We use three datasets for the experiments, as set out in Table~\ref{tab:datasets_summary}.
The first data set is for sentiment analysis.
This {\bf FPB} or Financial Phrase Bank Dataset consists of 4844 sentences or
phrases extracted from financial news, each of which is annotated with
a sentiment label: positive, negative, or neutral \citep{malo2014fpb, huggingface2021fpb}.
Each entry is labelled by six annotators and is categorised by the percentage of agreement among them.
This dataset has a consensus subset where all annotators agreed on the sentiment,
but we include all entries where the agreement is at least 50\%, giving good
diversity and complexity coverage for training.

\begin{table*}[t]
\centering
\caption{Summary of datasets used in the experiments.}
\label{tab:datasets_summary}
\begin{tabular}{p{1.1cm}c@{\hspace{5pt}}r@{\hspace{5pt}}r@{\hspace{10pt}}p{1.8cm}c@{\hspace{6pt}}p{3.2cm}p{1.7cm}}
\toprule
Dataset & & Size & Test & Task & \#labels & Label distr. & Avg. size \\ \midrule
FPB & & 4,844 & 969 & \makecell{Sentiment\\analysis} & 3 &
\makecell[l]{Negative: $\sim$59\% \\ Neutral: $\sim$28\% \\ Positive: $\sim$13\%}
& $\sim$18 words \\[0.4ex]
\hline
LIAR & & 12,836 & 1267 & \makecell{Fake news\\detection} & 6 &
\makecell[l]{True: 17\% \\ Mostly True: 21\% \\ Half True: 21\% \\ Mostly False: 20\% \\ False: 19\% \\ PoF: 2\%}
& $\sim$18 words \\[0.4ex]
\hline
BoolQ & & 27,427 & 1635 & \makecell{Binary\\Q\&A} & 2 & \makecell[l]{True: 59\% \\ False: 41\%} & $\sim$91 words \\

\bottomrule
\end{tabular}
\end{table*}

The second data set is for fake news detection.
Known as the {\bf LIAR} dataset, it consists of 12,836 labelled short statements
collected from PolitiFact.com, aimed at automated fact-checking research \citep{Wang2017acl}.
Each entry includes one of six labels indicating varying levels of truthfulness: `True', `Mostly True', `Half True', `Mostly False', `False', and `Pants on Fire'.

The third dataset is {\bf BoolQ} \citep{BoolQ2019naacl}. It contains factual questions in plain text and answer in Yes or No. We use this dataset for the Binary Q\&A task to compare performance of encoder-based models and decoder-based models.

Following standard experimental protocol,
in each case
the dataset is split into a training set, test set, and validation set.

\begin{table*}[t]
\centering
\caption{Summary of models adopted in fake news detection, sentiment analysis, and decoupling training tasks.}
\label{tab:models-summary}
\renewcommand{\arraystretch}{1}
\begin{tabular}{
    p{2.5cm} @{\hspace{6pt}} 
    p{2.8cm} @{\hspace{6pt}} 
    c @{\hspace{6pt}} 
    c @{\hspace{6pt}} 
    c @{\hspace{6pt}}
}
\toprule
Model type & Models & \makecell[c]{Fake news\\detection}
    & \makecell[c]{Sentiment\\analysis}
    & \makecell[c]{Binary\\Q\&A}  \\
\midrule

\multirow{4}{=}{Machine learning models} 
& \makecell[l]{SVM} & \checkmark & \checkmark & -- \\
& \makecell[l]{Logistic regression} & \checkmark & \checkmark & -- \\
& \makecell[l]{Na\"ive\\Bayesian} & \checkmark & \checkmark & -- \\
& \makecell[l]{Random forest} & \checkmark & \checkmark & -- \\ \midrule

\multirow{4}{=}{Deep learning models} 
& \makecell[l]{BERT} & \checkmark & \checkmark & \checkmark  \\
& \makecell[l]{RoBERTa} & \checkmark & \checkmark & \checkmark  \\
& \makecell[l]{XLNet} & \checkmark & \checkmark & \checkmark \\
& \makecell[l]{BART} & -- & -- & \checkmark \\
& \makecell[l]{T5} & -- & -- & \checkmark \\
\bottomrule
\end{tabular}
\vspace{2mm}

\textit{Note:} `\checkmark' indicates that the model has been
previously used for the task; `--' indicates it has not been applied.
\end{table*}

\subsubsection*{ML models}
In our experiments we have used
a variety of models including both classic ML approaches and transformer-based models.
For the former,
we use SVMs with a linear kernel,
logistic regression with the `saga' solver,
a na\"ive Bayesian model, and random forests.
In each case, we use the implementation from {\tt sci-kit learn} \citep{pedregosa2011jmlr}. 

For transformer-based models, we use Bidirectional Encoder Representations from Transformers
(BERT) \citep{devlin2019bert},
the baseline transformer-based model of many later derivatives;
Robustly Optimized BERT Pre-training Approach (RoBERTa) \citep{liu2019roberta},
which seeks to improve on BERT by removing the next-sentence prediction task, using
larger training datasets, training for more steps, and optimizing hyper-parameters;
XLNet \citep{yang2019xlnet}, a transformer-based model that combines
the strengths of autoregressive and auto-encoding methods by using permutation-based
language modelling to capture bidirectional context without relying on corrupting input tokens,
in contrast to BERT;
They are all trained with the same learning rate of $2^{-5}$
and 10 epochs with early stopping.

For the binary Q\&A task, we introduce two encoder-decoder-based models:
BART \citep{bart2020acl} and T5 \citep{t52020jmlr}.
These are of greater complexity than encoder-based models such as BERT and RoBERTa.
In more detail,
the BART model is a de-noising auto-encoder model for sequence-to-sequence generation tasks \citep{bart2020acl}.
The model adopts a similar but simplified architecture to a BERT model (for encoding), and a GPT model (for decoding). BART has shown competitive performance to RoBERTa on text generation tasks and also achieved new state-of-the-art on question answering task \citep{bart2020acl}. 

T5 is an encoder-decoder model trained for general text-to-text tasks \citep{t52020jmlr}.
By adopting the architecture of the original transformer model,
T5 takes textual embeddings as input and output texts in response.
In addition, it also facilitates a task-agnostic step on the input text so that it
can be fine-tuned for downstream tasks such as question answering and document summarisation.
In the reported performance on the binary Q\&A benchmark, T5-11B, a variant of~T5,
achieves higher performance scores than the previous best model, RoBERTa.
However, due to computational costs, we use the base version of T5 rather
than the superior variant. 

It could be argued that lack of fine tuning in our experiments 
is a confound.
However, such a claim would only be defensible if it is regarded
as reasonable to choose a seed post hoc based on validation or
test data, and if it is known that the same seed will be valid
in practical applications where such evaluation data is unavailable.
Indeed, given the observed lack of predictability of performance
of such methods from one data set to another, it is remarkable that
tuning on validation data
seems to consistently give strong results in reported work.

\section{Results}

We now present the results of the bootstrapping and random seed experiments on the three tasks.
For all three tasks, four evaluation metrics are reported: accuracy, precision, recall, and F1 score.

As noted earlier, for sentiment analysis and fake news detection we can compare
machine-learning models and deep-learning models, while the
binary Q\&A task can only be undertaken by deep-learning models; in that case
we compare encoder-based models with encoder-decoder-based models.

\subsection*{Bootstrapping}

Results for bootstrapping are in
Table~\ref{tab:bootstrap_all_std}
These results show that the measured results are in most cases reasonably accurate.
For F1-score, the standard deviations are consistently around or below 0.015, meaning
in broad terms that they are within a range of 0.06 with 95\% confidence.
Similar values are observed for the other measures, with the exception of precision on the
fake news task, where the standard deviation in one case is 0.067.

Also notable is that there is little distinction between the methods -- all approaches
have similar levels of accuracy.
These results are a reminder that bootstrapping is a valuable technique for understanding
the variability or uncertainty in score due to the composition of the test data.

\begin{table}[t]
\centering
\caption{Test-set bootstrapping results across tasks and models (mean $\pm$ standard deviation).
Smaller standard deviations indicate higher accuracy (or confidence) in the measured result.}
\label{tab:bootstrap_all_std}
\begin{tabular}{m{1.8cm}lc@{~~~}c@{~~~}c@{~~~}c}
\hline
{Task} & {Model} & {Accuracy} & {Precision} & {Recall} & {F1-score} \\
\hline
\multirow{7}{*}{Sentiment} 
 & SVM & 0.91 $\pm$ 0.007 & 0.88 $\pm$ 0.011 & 0.89 $\pm$ 0.010 & 0.89 $\pm$ 0.010 \\
 & Logistic r. & 0.90 $\pm$ 0.007 & 0.89 $\pm$ 0.011 & 0.83 $\pm$ 0.013 & 0.86 $\pm$ 0.010 \\
 & Naïve Bayes & 0.83 $\pm$ 0.009 & 0.82 $\pm$ 0.015 & 0.71 $\pm$ 0.015 & 0.75 $\pm$ 0.015 \\
 & Random f. & 0.96 $\pm$ 0.005 & 0.97 $\pm$ 0.005 & 0.95 $\pm$ 0.007 & 0.96 $\pm$ 0.006 \\
 & BERT    & 0.96 $\pm$ 0.005 & 0.95 $\pm$ 0.007 & 0.96 $\pm$ 0.007 & 0.96 $\pm$ 0.006 \\
 & RoBERTa & 0.82 $\pm$ 0.010 & 0.76 $\pm$ 0.014 & 0.81 $\pm$ 0.012 & 0.78 $\pm$ 0.013 \\
 & XLNet   & 0.64 $\pm$ 0.011 & 0.53 $\pm$ 0.014 & 0.54 $\pm$ 0.016 & 0.54 $\pm$ 0.014 \\
\hline
\multirow{7}{*}{Fake news} 
 & SVM & 0.66 $\pm$ 0.018 & 0.52 $\pm$ 0.033 & 0.51 $\pm$ 0.009 & 0.46 $\pm$ 0.014 \\
 & Logistic r. & 0.67 $\pm$ 0.018 & 0.55 $\pm$ 0.044 & 0.51 $\pm$ 0.008 & 0.44 $\pm$ 0.013 \\
 & Naïve Bayes & 0.68 $\pm$ 0.017 & 0.61 $\pm$ 0.067 & 0.51 $\pm$ 0.006 & 0.43 $\pm$ 0.011 \\
 & Random f. & 0.63 $\pm$ 0.017 & 0.52 $\pm$ 0.024 & 0.51 $\pm$ 0.014 & 0.49 $\pm$ 0.017 \\
 & BERT    & 0.67 $\pm$ 0.018 & 0.58 $\pm$ 0.036 & 0.52 $\pm$ 0.011 & 0.48 $\pm$ 0.016 \\
 & RoBERTa & 0.67 $\pm$ 0.017 & 0.34 $\pm$ 0.009 & 0.50 $\pm$ 0.000 & 0.40 $\pm$ 0.006 \\
 & XLNet   & 0.68 $\pm$ 0.017 & 0.68 $\pm$ 0.049 & 0.52 $\pm$ 0.007 & 0.46 $\pm$ 0.015 \\
\hline
\multirow{5}{*}{Binary QA} 
 & BERT     & 0.60 $\pm$ 0.014 & 0.60 $\pm$ 0.030 & 0.52 $\pm$ 0.005 & 0.43 $\pm$ 0.011 \\
 & RoBERTa  & 0.77 $\pm$ 0.010 & 0.77 $\pm$ 0.010 & 0.75 $\pm$ 0.010 & 0.75 $\pm$ 0.010 \\
 & XLNet    & 0.71 $\pm$ 0.011 & 0.70 $\pm$ 0.011 & 0.68 $\pm$ 0.010 & 0.67 $\pm$ 0.011 \\
 & BART     & 0.60 $\pm$ 0.014 & 0.64 $\pm$ 0.042 & 0.51 $\pm$ 0.004 & 0.40 $\pm$ 0.009 \\
 & T5       & 0.69 $\pm$ 0.012 & 0.68 $\pm$ 0.012 & 0.65 $\pm$ 0.011 & 0.66 $\pm$ 0.012 \\
\hline
\end{tabular}
\end{table}

\subsection*{Random seeds}
We now examine the ML models as seeds are randomly varied, considering each task
in turn.

\subsubsection*{Sentiment analysis.}

\begin{table}[t]
\caption{Performance on the FPB sentiment-analysis dataset.
Results shown are mean and standard deviation across 100 random seeds.}
\label{tab:SA_rand}
\centering
\begin{tabular}{lcc@{~~~}c@{~~~}c@{~~~}c}
\toprule
Model && Accuracy & Precision & Recall & F1-Score \\
\midrule
SVM && $0.91 \pm 0.000$ & $0.88 \pm 0.000$ & $0.89 \pm 0.000$ & $0.89 \pm 0.000$ \\
Logistic regression && $0.90 \pm 0.000$ & $0.89 \pm 0.000$ & $0.83 \pm 0.000$ & $0.85 \pm 0.000$ \\
Naïve Bayes && $0.82 \pm 0.000$ & $0.82 \pm 0.000$ & $0.70 \pm 0.000$ & $0.74 \pm 0.000$ \\
Random forest && $0.96 \pm 0.001$ & $0.97 \pm 0.002$ & $0.95 \pm 0.001$ & $0.96 \pm 0.001$ \\
BERT && $0.88 \pm 0.065$ & $0.83 \pm 0.091$ & $0.84 \pm 0.115$ & $0.83 \pm 0.111$ \\
RoBERTa && $0.83 \pm 0.083$ & $0.76 \pm 0.169$ & $0.79 \pm 0.169$ & $0.76 \pm 0.179$ \\
XLNet && $0.78 \pm 0.100$ & $0.68 \pm 0.194$ & $0.69 \pm 0.202$ & $0.67 \pm 0.209$ \\
\bottomrule
\end{tabular}
\end{table}

Results of the random seed experiment on the sentiment analysis task are
in Table~\ref{tab:SA_rand}.
As expected, the statistical models show little or no
variance over 100 runs with different random seeds.
In contrast, the deep-learning models suffer from severe performance instability.
The F1-score of XLNet ranges widely, with a standard deviation of over 0.2, meaning, statistically,
that the actual score is essentially unknown -- likely to be between 0.4 and 1.0 but no
more specific than that.
Similar behaviour is observed for RoBERTa, where on this information
at best the F1-score could be asserted to lie between
0.6 and 1.0, and to a lesser extent BERT, with the range 0.7 to~1.0;
and also is observed for these methods on the other metrics.
These results demonstrate that seeds are a huge contributor to model performance, far outweighing
those due to the minor model alterations reported in many papers.

\begin{figure}[t]
      \centering
  \begin{subfigure}[t]{0.48\linewidth}
    \includegraphics[width=\linewidth]{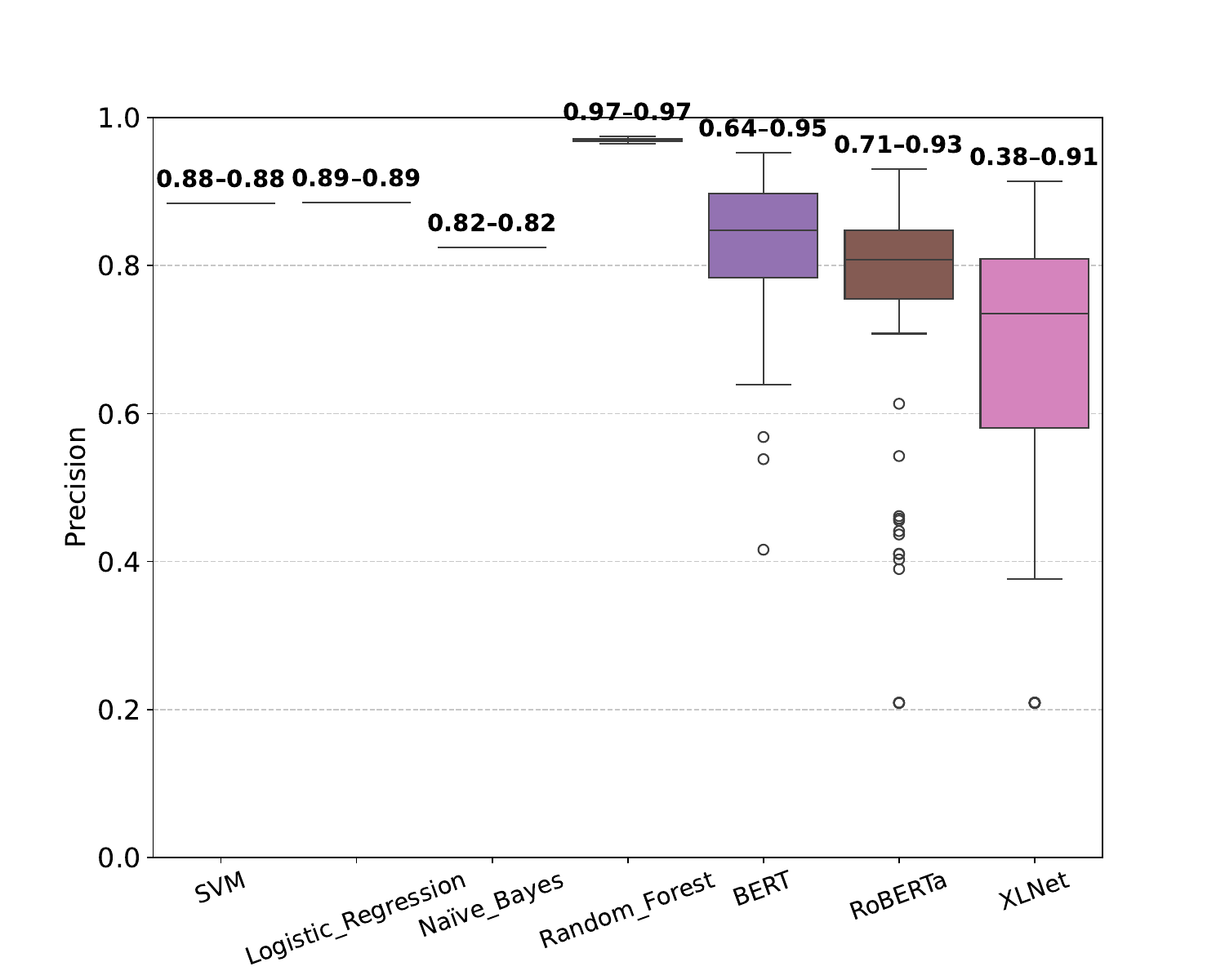}
    \caption{Precision}
    \label{fig:SA_precision_rand_a}
  \end{subfigure}
  \hfill
  \begin{subfigure}[t]{0.48\linewidth}
    \includegraphics[width=\linewidth]{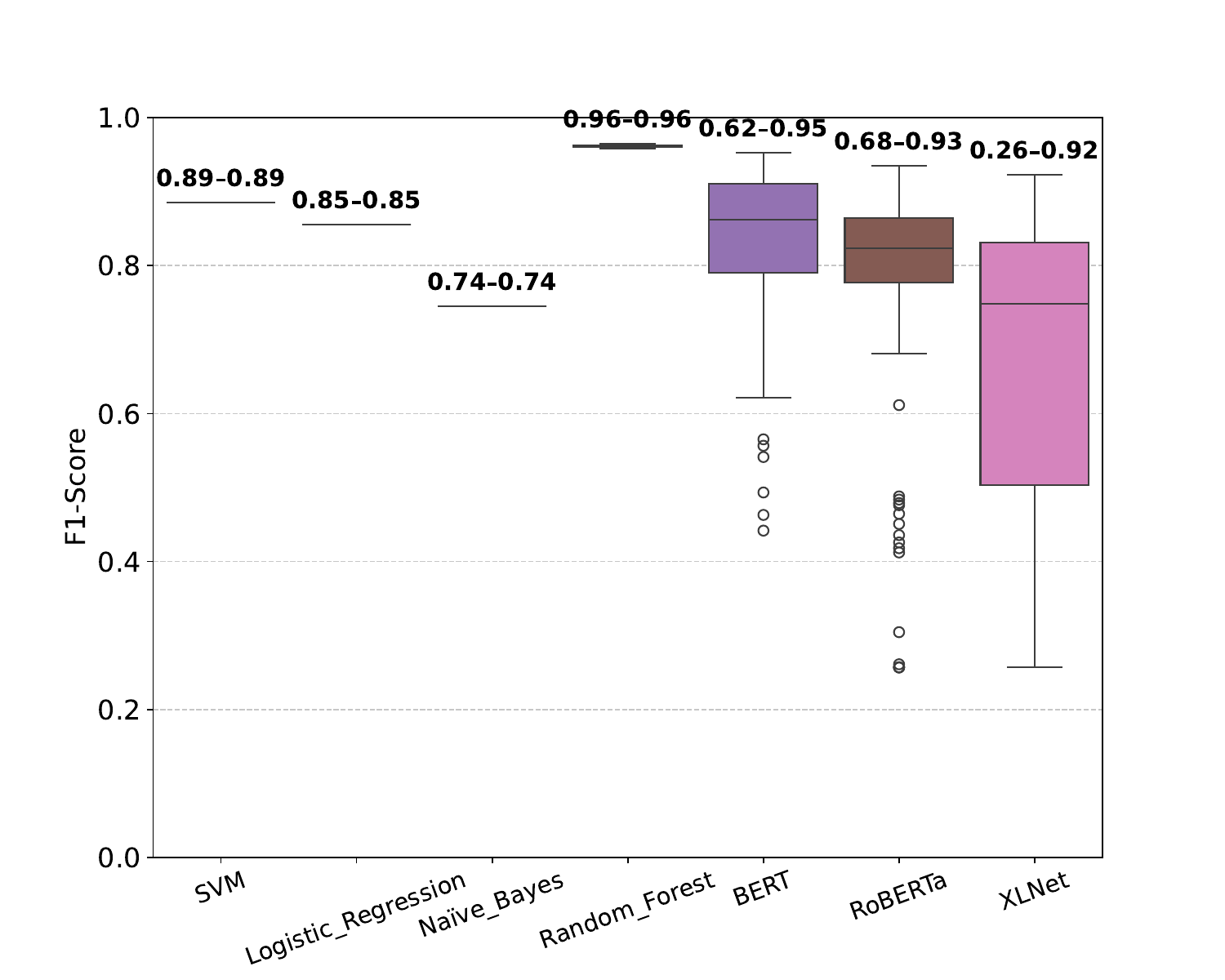}
    \caption{F1-score}
    \label{fig:SA_f1_rand_b}
  \end{subfigure}
  \caption{Precision (a) and F1-scores (b) over 100 random seeds on the
  FPB sentiment analysis dataset.
  Boxes show the interquartile range.
  Ranges excluding outliers beyond 1.5$\times$ the inter-quartile range are annotated for each model.}
  \label{fig:SA_precision_rand}
\end{figure}

These averages and deviations arise in complex ways for the different models; some are more nearly
normally distributed than others.
To demonstrate this,
we present precision and F1-score of all seven models in Figure~\ref{fig:SA_precision_rand}.
The performance of each model is represented with a box plot, and outliers at 1.5 times
the inter-quartile range are shown numerically.

The bootstrapping results, with standard deviations of around a tenth of that observed
under random seeds, demonstrate that these large variations in score are not arising in some
way from the test data.
These variations are due to instability in the methods being tested.

\subsubsection*{Fake news detection.}

\begin{table}[t]
\caption{Performance on the LIAR fake news detection dataset.
Results shown are mean and standard deviation across 100 random seeds. }
\centering
\begin{tabular}{lcc@{~~~}c@{~~~}c@{~~~}c}
\toprule
Model && Accuracy & Precision & Recall & F1-score \\
\midrule
SVM && $0.66\pm0.000$ & $0.52\pm0.000$ & $0.51\pm0.000$ & $0.46\pm0.000$ \\
Logistic regression && $0.67\pm0.000$ & $0.55\pm0.000$ & $0.51\pm0.000$ & $0.44\pm0.000$ \\
Naïve Bayes && $0.68\pm0.000$ & $0.61\pm0.000$ & $0.51\pm0.000$ & $0.43\pm0.000$ \\
Random forest && $0.64\pm0.007$ & $0.54\pm0.012$ & $0.52\pm0.007$ & $0.50\pm0.009$ \\
BERT && $0.68\pm0.009$ & $0.63\pm0.163$ & $0.52\pm0.016$ & $0.46\pm0.042$ \\
RoBERTa && $0.52\pm0.018$ & $0.45\pm0.135$ & $0.52\pm0.018$ & $0.47\pm0.091$ \\
XLNet && $0.51\pm0.011$ & $0.46\pm0.106$ & $0.51\pm0.011$ & $0.47\pm0.073$ \\
\bottomrule
\end{tabular}
\label{tab:fake_news_seed}
\end{table}

Results on LIAR are shown in Table~\ref{tab:fake_news_seed}.
The same pattern is repeated, albeit with reduced standard deviations.
Variation for the statistical models is small or absent.
For the transformer models, however, the variation is large, and
much greater than under bootstrapping; here, the worst results are for precision,
but we can only conclude that BERT's F1-score is between 0.35 and 0.55, say, while
RoBERTa's is between 0.3 and 0.65.
These are not results on which it can be argued that one method is superior to another.

\subsubsection*{Binary Q\&A.}

Experiments on the binary Q\&A task are only with transformer models.
During the training of deep-learning models, to ensure fair comparison for all models
we used a consistent learning rate of $10^{-5}$, batch size of 16, and 10 epochs; this
last choice was determined based on multiple trials of running the models
and ensured a visible decrease of training loss and evaluation loss while avoiding under-fitting.
We avoided further fine-tuning of each model to the data, as this relative performance was not
a factor we were measuring.

Results are in Table~\ref{tab:bianry_QA_rand}.
The first four models all exhibit huge variation, to the extent that their scores are effectively
unknown, as in the previous experiments.
Intriguingly, T5 was much more stable, with variation similar to that under bootstrapping.
While there were large numbers of seeds where the other methods had higher performance, their
huge score uncertainties meant that no inference can be drawn from these instances.

\begin{table}[t]
\caption{Performance on the BoolQ Binary Q\&A dataset.
Results are mean and standard deviation across 100 random seeds.}
\label{tab:bianry_QA_rand}
\centering
\begin{tabular}{lcc@{~~~}c@{~~~}c@{~~~}c}
\toprule
Model && Accuracy & Precision & Recall & F1-Score \\
\midrule
BERT && $0.64 \pm 0.040$ & $0.61 \pm 0.104$ & $0.60 \pm 0.071$ & $0.56 \pm 0.117$ \\
RoBERTa && $0.64 \pm 0.060$ & $0.48 \pm 0.198$ & $0.57 \pm 0.093$ & $0.50 \pm 0.151$ \\
XLNet && $0.61 \pm 0.022$ & $0.48 \pm 0.182$ & $0.53 \pm 0.038$ & $0.43 \pm 0.083$ \\
BART && $0.70 \pm 0.109$ & $0.57 \pm 0.248$ & $0.65 \pm 0.147$ & $0.59 \pm 0.210$ \\
T5 && $0.67 \pm 0.016$ & $0.66 \pm 0.018$ & $0.64 \pm 0.019$ & $0.64 \pm 0.020$ \\
\bottomrule
\end{tabular}
\end{table}

\section{Discussion and conclusions}

We have explored variation in performance of ML methods due to choice of random seed,
across three IR classification tasks.
Our key finding is that for transformer models
this variation can be dramatic -- so great that the results are
effectively unknown.

In the cited papers on which our study is based, there are
no cases in which the effect of varying seeds is reported, and
indeed many researchers report results with high confidence.
One of these papers reports that RoBERTa outperforms XLNet by 0.0132, and another makes the
same claim of outperformance, by~0.008;
and another claims that RoBERTa outperforms BERT by 0.0173.
There is no mention of confidence or of the validity
of reporting to four decimal places.

More broadly we have observed such claims widely in the literature -- and for this
reason we do not give citations for the above, to avoid singling out individuals
for what is a widespread behaviour.
These claims are almost never
accompanied by caveats or exploration of the impact of choice of seed,
or acknowledgement of the extent to which seeds explored to
tune the method.
Our conclusion is simple: without such analysis, these claims are not valid.

Our results may explain another phenomenon, however: that different experiments
give different relativities between the methods.
We have observed that the uncertainties in measurement are far greater than the
average differences in performance.
It is thus unsurprising that results have been inconsistent.



These results show the need to report performance variance over multiple random seeds;
the complexity of infrastructure and the indeterministic nature of the training process bring high uncertainty to model robustness.
It seems likely that some researchers are indeed trying different random seeds until good results are
obtained, under the practice of `fine tuning'.
Even if this practice were defensible -- a position we do not agree with -- the extent of variation
is a weakness, and should be acknowledged.
Such practices raise critical questions about the 
reliability and interpretability of published findings, and highlight 
the need for systematic evaluation of performance variation in 
transformer-based methods.

%
%
%
\bibliographystyle{plainnat}
\bibliography{clean_reference}

\end{document}